\documentstyle[12pt,qqa4lart]{article}
\input tcilatex

\QQQ{Language}{
American English
}

\begin{document}

\author{Lu-Ming Duan and Guang-Can Guo\thanks{%
Electronic address: gcguo@sunlx06.nsc.ustc.edu.cn} \\
Department of Physics and Nonlinear Science Center,\\
University of Science and Technology of China,\\
Hefei 230026, People's Republic of China}
\title{Probabilistic cloning and identification of linearly independent quantum
states}
\date{}
\maketitle

\begin{abstract}
\baselineskip 24pt We construct a probabilistic quantum cloning machine by a
general unitary-reduction operation. With a postselection of the measurement
results, the machine yields faithful copies of the input states. It is shown
that the states secretly chosen from a certain set $\$=\left\{ \left| \Psi
_1\right\rangle ,\left| \Psi _2\right\rangle ,\cdots ,\left| \Psi
_n\right\rangle \right\} $ can be probabilistically cloned if and only if $%
\left| \Psi _1\right\rangle $, $\left| \Psi _2\right\rangle $, $\cdots ,$
and $\left| \Psi _n\right\rangle $ are linearly-independent. We derive the
best possible cloning efficiencies. Probabilistic cloning has close
connection with the problem of identification of a set of states, which is a
type of $n+1$ outcome measurement on $n$ linearly independent states. The
optimal efficiencies for this type of measurement are obtained.\\

\end{abstract}

\newpage\baselineskip 24pt In quantum mechanics, a combination of unitary
evolution together with measurements often yields interesting results, such
as the quantum programming [1], the purification of entanglements [2], and
the teleportation [3] and preparation [4] of quantum states. Here, we
discuss such a combination in the field of quantum cloning. With the
development of quantum information theory, quantum cloning has become a
field of growing interest [5-19]. We should discriminate two kinds of
quantum cloning. If a cloning machine performs merely unitary operations, it
is called deterministic cloning, since unitary evolution is deterministic.
On the other hand, if a cloning machine performs measurements as well as
unitary operations, with a postselection of the measurement results, it is
called probabilistic cloning, since the desired copies are produced only
with certain probabilities. There are two different types of statements for
the quantum no-cloning theorem. The first [5] asserts that an arbitrary
unknown state can not be cloned, whether deterministically or
probabilistically, since the linearity of quantum operations forbids such a
replication; and the second [7-9] states that deterministic cloning of
nonorthogonal states is impossible because of the unitarity of the
evolution. The quantum no-cloning theorem does not rule out the possibility
of probabilistic cloning of nonorthogonal states. In fact, two nonorthogonal
states can indeed be probabilistically cloned [18]. Then, what property
characterizes the set of states able to be probabilistically cloned? In this
paper, we show that the states secretly chosen from a certain set can be
probabilistically cloned if and only if they are linearly independent. We
also derive the best possible cloning efficiencies.

It is appropriate to emphasize differences between the probabilistic cloning
and the inaccurate quantum copying more extensively discussed in recent
literatures [10-17]. The inaccurate copying process is unitary and thus
deterministic. For nonorthogonal states, the state fidelity can never attain
1. Arbitrary unknown states are able to be inaccurately copied. In contrast,
only linearly independent states can be probabilistically cloned. The
probabilistic cloning machine yields faithful copies of the input states
with certain non-zero probabilities of success. For this machine, the
inaccurate copies are discarded.

The $1\rightarrow 2$ cloning machine produces two copies of the input state.
Similarly, we may consider $1\rightarrow m$, and even $1\rightarrow \infty $
cloning machines. We will show that the probabilistic cloning is closely
related to the problem of identification of a set of states. The
identification measurement differs from Helstrom's minimal error probability
decision [20]. It is in fact an $n+1$ outcome measurement on $n$ possible
input states $\left| \Psi _1\right\rangle $, $\left| \Psi _2\right\rangle $, 
$\cdots ,$ and $\left| \Psi _n\right\rangle $. The outcome $i$ $%
(i=1,2,\cdots ,$ or $n)$ corresponds that the state is definitely $\left|
\Psi _i\right\rangle $, whereas the $n+1$ outcome is ''failure'',
corresponding the case that we cannot identify what the state really is form
the measurement result. The Helstrom measurement does not determine what the
state really is. It is succeeded by a guess and the minimal error
probability is required in the guess.

We start by showing that only linearly independent states can be
probabilistically cloned. This is the following theorem.

{\it Theorem 1.} The states secretly chosen from the set $\$=\left\{ \left|
\Psi _1\right\rangle ,\left| \Psi _2\right\rangle ,\cdots ,\left| \Psi
_n\right\rangle \right\} $ can be probabilistically cloned by a general
unitary-reduction operation if and only if $\left| \Psi _1\right\rangle $, $%
\left| \Psi _2\right\rangle $, $\cdots ,$ and $\left| \Psi _n\right\rangle $
are linearly independent.

{\it Proof.} Any operation in quantum mechanics can be represented by a
unitary evolution together with a measurement [21]. To get faithful copies
of the pure input states $\left| \Psi _i\right\rangle $, the output states
of the cloning machine are also pure. This requires that the measurement in
the cloning machine should be performed with a postselection of the
measurement results. A measurement with a postselection of the measurement
results is described by a projection operator, and like the unitary
evolution, it is linear in the state vector. Hence, similar to the original
proof of the no-cloning theorem [5], this linearity forbids faithful cloning
of linearly dependent quantum states, whether in a deterministic or in a
probabilistic fashion. Our task remains to prove the converse, that is, to
show that if $\left| \Psi _1\right\rangle $, $\left| \Psi _2\right\rangle $, 
$\cdots $, and $\left| \Psi _n\right\rangle $ are $n$ linearly-independent
states of a system A, there exist a unitary operation $U$ and a measurement $%
M$, which together yield the following evolution 
\begin{equation}
\label{1}\left| \Psi _i\right\rangle \left| \Sigma \right\rangle \stackrel{%
U+M}{\longrightarrow }\left| \Psi _i\right\rangle \left| \Psi
_i\right\rangle ,\text{ }\left( i=1,2,\cdots ,n\right) , 
\end{equation}
where $\left| \Sigma \right\rangle $ is the input state of an ancillary
system B. Systems A and B each have an $N$-dimensional Hilbert space with $%
N\geq n$.

To prove the above statement, we introduce a probe P with an $n_p-$%
dimensional Hilbert space, where $n_p\geq n+1$. Suppose $\left|
P_0\right\rangle $, $\left| P_1\right\rangle $, $\cdots $, and $\left|
P_n\right\rangle $ are $n+1$ orthonormal states of the probe P. If there
exists a unitary operator $U$ to make 
\begin{equation}
\label{2}U\left( \left| \Psi _i\right\rangle \left| \Sigma \right\rangle
\left| P_0\right\rangle \right) =\sqrt{\gamma _i}\left| \Psi _i\right\rangle
\left| \Psi _i\right\rangle \left| P_0\right\rangle +\stackrel{n}{%
\stackunder{j=1}{\sum }}c_{ij}\left| \Phi _{AB}^{\left( j\right)
}\right\rangle \left| P_j\right\rangle ,\text{ }\left( i=1,2,\cdots
,n\right) , 
\end{equation}
where $\left| \Phi _{AB}^{\left( 1\right) }\right\rangle $, $\left| \Phi
_{AB}^{\left( 2\right) }\right\rangle $, $\cdots $, and $\left| \Phi
_{AB}^{\left( n\right) }\right\rangle $ are $n$ normalized states of the
composite system AB (not generally orthogonal), we measure the probe P after
the evolution. The cloning attempt has succeeded and the output state of the
system AB is kept if and only if the measurement outcome of the probe is $%
P_0 $. With probability $\gamma _i$ of success, this measurement projects
the composite system AB into the replicated state $\left| \Psi
_i\right\rangle \left| \Psi _i\right\rangle $, where $i=0$, $1$, $\cdots ,$
or $n$. The parameters $\gamma _i$ are called the cloning efficiencies. For
any input state $\left| \Psi _i\right\rangle $, the probabilistic cloning
machine should succeed with a non-zero probability. This requires that all
the $\gamma _i$ be positive real numbers. Therefore, the evolution (1) can
be realized in physics if Eq. (2) holds with positive efficiencies. To prove
existence of the unitary evolution described by Eq. (2), we notice the
following fact.

{\it Lemma 1.} If two sets of states $\left| \phi _1\right\rangle $, $\left|
\phi _2\right\rangle $, $\cdots $, $\left| \phi _n\right\rangle $, and $%
\left| \widetilde{\phi }_1\right\rangle $, $\left| \widetilde{\phi }%
_2\right\rangle $, $\cdots ,$ $\left| \widetilde{\phi }_n\right\rangle $
satisfy the condition 
\begin{equation}
\label{3}\left\langle \phi _i|\phi _j\right\rangle =\left\langle \widetilde{%
\phi }_i|\widetilde{\phi }_j\right\rangle ,\text{ }\left( i=1,2,\cdots ,n;%
\text{ }j=1,2,\cdots ,n\right) , 
\end{equation}
there exists a unitary operator $U$ to make $U\left| \phi _i\right\rangle
=\left| \widetilde{\phi }_i\right\rangle ,$ $\left( i=1,2,\cdots ,n\right) $.

The $n\times n$ inter-inner-products of Eq. (2) yield the matrix equation 
\begin{equation}
\label{4}X^{\left( 1\right) }=\sqrt{\Gamma }X^{\left( 2\right) }\sqrt{\Gamma 
}^{+}+CC^{+}, 
\end{equation}
where the $n\times n$ matrixes $C=\left[ c_{ij}\right] $, $X^{\left(
1\right) }=\left[ \left\langle \Psi _i|\Psi _j\right\rangle \right] $, and $%
X^{\left( 2\right) }=\left[ \left\langle \Psi _i|\Psi _j\right\rangle
^2\right] $. The diagonal efficiency matrix $\Gamma $ is defined by $\Gamma
=diag\left( \gamma _1,\gamma _2,\cdots ,\gamma _n\right) $, hence $\sqrt{%
\Gamma }=\sqrt{\Gamma }^{+}=diag\left( \sqrt{\gamma _1},\sqrt{\gamma _2}%
,\cdots ,\sqrt{\gamma _n}\right) $. Lemma 1 shows that if Eq. (4) is
satisfied with a diagonal positive-definite matrix $\Gamma $, the unitary
evolution (2) can be realized in physics.

To prove that there is a diagonal positive-definite matrix $\Gamma $ to
satisfy Eq. (4), first we show that the matrix $X^{\left( 1\right) }$ is
positive-definite. This is the following lemma.

{\it Lemma 2.} If $n$ states $\left| \Psi _1\right\rangle $, $\left| \Psi
_2\right\rangle $, $\cdots ,$ and $\left| \Psi _n\right\rangle $ are
linearly independent, the matrix $X^{\left( 1\right) }=\left[ \left\langle
\Psi _i|\Psi _j\right\rangle \right] $ is positive-definite.

{\it Proof of Lemma 2.} For an arbitrary $n$-vector $B=\left( b_1,b_2,\cdots
,b_n\right) ^T$, the quadratic form $B^{+}X^{\left( 1\right) }B$ can be
expressed as 
\begin{equation}
\label{5}B^{+}X^{\left( 1\right) }B=\left\langle \Psi _T|\Psi
_T\right\rangle =\left\| \left| \Psi _T\right\rangle \right\| ^2, 
\end{equation}
where 
\begin{equation}
\label{6}\left| \Psi _T\right\rangle =b_1\left| \Psi _1\right\rangle
+b_2\left| \Psi _2\right\rangle +\cdots +b_n\left| \Psi _n\right\rangle . 
\end{equation}
Since the states $\left| \Psi _1\right\rangle $, $\left| \Psi
_2\right\rangle $, $\cdots ,$ and $\left| \Psi _n\right\rangle $ are
linearly independent, the summation state $\left| \Psi _T\right\rangle $
does not reduce to zero for any $n$-vector $B$ and its norm is thus always
positive. By definition, the matrix $X^{\left( 1\right) }$ is
positive-definite.

Since $X^{\left( 1\right) }$ is positive-definite, from continuity, for
small enough but positive $\gamma _i$, the matrix $X^{\left( 1\right) }-%
\sqrt{\Gamma }X^{\left( 2\right) }\sqrt{\Gamma }^{+}$ is also
positive-definite. So the Hermitian matrix $X^{\left( 1\right) }-\sqrt{%
\Gamma }X^{\left( 2\right) }\sqrt{\Gamma }^{+}$ is able to be diagonalized
by a unitary matrix $V$ as follows 
\begin{equation}
\label{7}V^{+}\left( X^{\left( 1\right) }-\sqrt{\Gamma }X^{\left( 2\right) }%
\sqrt{\Gamma }^{+}\right) V=\text{diag}\left( m_1,m_2,\cdots ,m_n\right) , 
\end{equation}
where all the eigenvalues $m_1,$ $m_2,$ $\cdots ,$ and $m_n$ are positive
real numbers. In Eq. (4), the matrix $C$ can be chosen as 
\begin{equation}
\label{8}C=V\text{diag}\left( \sqrt{m_1},\sqrt{m_2},\cdots ,\sqrt{m_n}%
\right) V^{+}. 
\end{equation}
Eq. (4) is thus satisfied with a diagonal positive-definite efficiency
matrix $\Gamma $. This completes the proof of theorem 1.

In the above proof, the condition of linearly independence of the $n$ states 
$\left| \Psi _1\right\rangle $, $\left| \Psi _2\right\rangle $, $\cdots ,$
and $\left| \Psi _n\right\rangle $ plays an essential role. If $\left| \Psi
_1\right\rangle $, $\left| \Psi _2\right\rangle $, $\cdots ,$ and $\left|
\Psi _n\right\rangle $ are linearly dependent, there exists an $n$-vector $B$
to make $B^{+}X^{\left( 1\right) }B=0,$ and the matrix $X^{\left( 1\right) }$
is therefore only positive-semidefinite. With a diagonal positive-definite
matrix $\Gamma $, in general, $X^{\left( 1\right) }-\sqrt{\Gamma }X^{\left(
2\right) }\sqrt{\Gamma }^{+}$ is no longer a positive-semidefinite matrix.
But the matrix $CC^{+}$ is positive--semidefinite. So Eq. (4) cannot be
satisfied. This shows in an alternative way that $n$ linearly dependent
states $\left| \Psi _1\right\rangle $, $\left| \Psi _2\right\rangle $, $%
\cdots ,$ and $\left| \Psi _n\right\rangle $ cannot be probabilistically
cloned by any unitary-reduction operation.

Deterministic cloning can be regarded as a special case of the probabilistic
cloning, with all the cloning efficiencies $\gamma _i=1$. For nonorthogonal
states, at least some of the $\gamma _i$ are less than $1$ . If all the $%
\gamma _i=1$, i.e., $\Gamma =I_n,$ Eq. (4) reduces to $X^{\left( 1\right)
}=X^{\left( 2\right) }.$ This is possible if and only if the states $\left|
\Psi _1\right\rangle $, $\left| \Psi _2\right\rangle $, $\cdots ,$ and $%
\left| \Psi _n\right\rangle $ are orthogonal to each other. Hence,
non-orthogonal states can not be deterministically cloned by the same
machine. This is a well-known result and it has important implications in
quantum cryptography [22-25].

In the following, we derive the best possible efficiencies $\gamma _i$ able
to be attained by a probabilistic cloning machine. A general unitary
evolution of the system ABP can be decomposed as 
\begin{equation}
\label{9}U\left( \left| \Psi _i\right\rangle \left| \Sigma \right\rangle
\left| P_0\right\rangle \right) =\sqrt{\gamma _i}\left| \Psi _i\right\rangle
\left| \Psi _i\right\rangle \left| P^{\left( i\right) }\right\rangle +\sqrt{%
1-\gamma _i}\left| \Phi _{ABP}^{\left( i\right) }\right\rangle ,\text{ }%
\left( i=1,2,\cdots ,n\right) , 
\end{equation}
where $\left| P_0\right\rangle $ and $\left| P^{\left( i\right)
}\right\rangle $ are normalized states of the probe P (not generally
orthogonal) and $\left| \Phi _{ABP}^{\left( 1\right) }\right\rangle $, $%
\left| \Phi _{ABP}^{\left( 2\right) }\right\rangle $, $\cdots $, and $\left|
\Phi _{ABP}^{\left( n\right) }\right\rangle $ are $n$ normalized states of
the composite system ABP (not generally orthogonal). Without loss of
generality, in Eq. (9) the coefficients before the states $\left| \Psi
_i\right\rangle \left| \Psi _i\right\rangle \left| P^{\left( i\right)
}\right\rangle $ and $\left| \Phi _{ABP}^{\left( i\right) }\right\rangle $
are assumed to be positive real numbers. Obviously, Eq. (2) is a special
case of Eq. (9) with $\left| P_0\right\rangle $ and all $\left| P^{\left(
i\right) }\right\rangle $ being the same state and $\left| \Phi
_{ABP}^{\left( i\right) }\right\rangle $ having a special decomposition. We
denote the subspace spanned by the states $\left| P^{\left( 1\right)
}\right\rangle $, $\left| P^{\left( 2\right) }\right\rangle $, $\cdots $,
and $\left| P^{\left( n\right) }\right\rangle $ by the symbol $S_0$. During
the cloning process, after the unitary evolution a measurement of the probe
with a postselection of the measurement results projects its state into the
subspace $S_0$. After this projection, the state of the system AB should be $%
\left| \Psi _i\right\rangle \left| \Psi _i\right\rangle $, so all the states 
$\left| \Phi _{ABP}^{\left( i\right) }\right\rangle $ ought to lie in a
space orthogonal to $S_0$. This requires that $\left| \Phi _{ABP}^{\left(
j\right) }\right\rangle $ be annihilated by the projection operator $\left|
P^{\left( i\right) }\right\rangle \left\langle P^{\left( i\right) }\right| $
for any $i$ and $j$, i.e., 
\begin{equation}
\label{10}\left| P^{\left( i\right) }\right\rangle \left\langle P^{\left(
i\right) }\right| \left| \Phi _{ABP}^{\left( j\right) }\right\rangle =0,%
\text{ }\left( i=1,2,\cdots ,n;\text{ }j=1,2,\cdots ,n\right) . 
\end{equation}
Under the condition (10), inter-inner-products of Eq. (9) yield the
following matrix equation 
\begin{equation}
\label{11}X^{\left( 1\right) }=\sqrt{\Gamma }X_P^{\left( 2\right) }\sqrt{%
\Gamma }^{+}+\sqrt{I_n-\Gamma }Y\sqrt{I_n-\Gamma }^{+}, 
\end{equation}
where the $n\times n$ matrixes $Y=\left[ <\Phi _{ABP}^{\left( i\right)
}|\Phi _{ABP}^{\left( j\right) }>\right] $ and $X_P^{\left( 2\right)
}=\left[ \left\langle \Psi _i|\Psi _j\right\rangle ^2\left\langle P^{\left(
i\right) }|P^{\left( j\right) }\right\rangle \right] $, and $I_n$ is the $%
n\times n$ unit matrix. Following the proof of lemma 2, $Y$, and thus $\sqrt{%
I_n-\Gamma }Y\sqrt{I_n-\Gamma }^{+}$, are positive-semidefinite matrixes, so 
$X^{\left( 1\right) }-\sqrt{\Gamma }X_P^{\left( 2\right) }\sqrt{\Gamma }^{+}$
should also be positive-semidefinite. On the other hand, if $X^{\left(
1\right) }-\sqrt{\Gamma }X_P^{\left( 2\right) }\sqrt{\Gamma }^{+}$ is a
positive-semidefinite matrix, following the proof of theorem 1, Eq. (11) can
be satisfied with a special choice of $\left| \Phi _{ABP}^{\left( i\right)
}\right\rangle $, and then lemma 1 shows that the states $\left| \Psi
_1\right\rangle $, $\left| \Psi _2\right\rangle $, $\cdots ,$ and $\left|
\Psi _n\right\rangle $ are able to be probabilistically cloned. We thus get
the following theorem.

{\it Theorem 2.} The states $\left| \Psi _1\right\rangle $, $\left| \Psi
_2\right\rangle $, $\cdots ,$ and $\left| \Psi _n\right\rangle $ can be
probabilistically cloned with a diagonal efficiency matrix $\Gamma $ if and
only if the matrix $X^{\left( 1\right) }-\sqrt{\Gamma }X_P^{\left( 2\right) }%
\sqrt{\Gamma }^{+}$ is positive-semidefinite.

The semi-positivity of the matrix $X^{\left( 1\right) }-\sqrt{\Gamma }%
X_P^{\left( 2\right) }\sqrt{\Gamma }^{+}$ gives a series of inequalities
about the efficiencies $\gamma _i$. The best possible cloning efficiencies $%
\gamma _i$ are obtained by solving these inequalities and then taking
maximum over all possible choices of the normalized states $\left| P^{\left(
i\right) }\right\rangle $. For example, if there are only two states $\left|
\Psi _1\right\rangle $ and $\left| \Psi _2\right\rangle $, theorem 2 shows
that the cloning efficiencies $\gamma _1$ and $\gamma _2$ satisfy 
\begin{equation}
\label{12}\frac{\gamma _1+\gamma _2}2\leq \stackunder{\left| P^{\left(
i\right) }\right\rangle }{\max }\frac{1-\left| <\Psi _1|\Psi _2>\right| }{%
1-\left| <\Psi _1|\Psi _2>\right| ^2\left| \left\langle P^{\left( 1\right)
}|P^{\left( 2\right) }\right\rangle \right| }=\frac 1{1+\left| <\Psi _1|\Psi
_2>\right| }, 
\end{equation}
where we assumed $\left| <\Psi _1|\Psi _2>\right| \neq 1$. The equality in
Eq. (12) holds if and only if $\gamma _1=\gamma _2$ and $\left\langle
P^{\left( 1\right) }|P^{\left( 2\right) }\right\rangle <\Psi _1|\Psi
_2>=\left| <\Psi _1|\Psi _2>\right| $. The best possible efficiencies
obtained from theorem2 depend on inner-products of the input states. This is
a natural result since probabilistic cloning is possible only for a known
set of states. Theorem 2 is a basic result in determining the best possible
cloning efficiencies.

The analysis of the $1\rightarrow 2$ probabilistic cloning can be directly
extended to include the $1\rightarrow m$ probabilistic cloning. The
extension is straightforward, and we omit its proof. The result is

{\it Theorem 3.} The states $\left| \Psi _1\right\rangle $, $\left| \Psi
_2\right\rangle $, $\cdots ,$ and $\left| \Psi _n\right\rangle $ can be
probabilistically replicated into $m$ faithful copies with a diagonal
efficiency matrix $\Gamma $ if and only if the matrix $X^{\left( 1\right) }-%
\sqrt{\Gamma }X_P^{\left( m\right) }\sqrt{\Gamma }^{+}$ is
positive-semidefinite.

The matrix $X_P^{\left( m\right) }$ in theorem 3 is defined by $X_P^{\left(
m\right) }=\left[ \left\langle \Psi _i|\Psi _j\right\rangle ^m\left\langle
P^{\left( i\right) }|P^{\left( j\right) }\right\rangle \right] $. The $%
1\rightarrow \infty $ probabilistic cloning is of special interest. It is
closely related to the problem of identification of the states $\left| \Psi
_1\right\rangle $, $\left| \Psi _2\right\rangle $, $\cdots ,$ and $\left|
\Psi _n\right\rangle $. On the one hand, if we have infinitely many faithful
copies of the input state, the state can be definitely determined. On the
other hand, if the input state is definitely determined, we can generate
infinitely many faithful copies. The best possible efficiencies for the $%
1\rightarrow \infty $ probabilistic cloning is determined by the
semi-positivity of the matrix $X^{\left( 1\right) }-\sqrt{\Gamma }%
X_P^{\left( \infty \right) }\sqrt{\Gamma }^{+}=X^{\left( 1\right) }-\Gamma $%
, where we assumed $\left| \left\langle \Psi _i|\Psi _j\right\rangle \right|
<1$ for $i\neq j$. Are these the optimal efficiencies for the identification
measurement on the states $\left| \Psi _1\right\rangle $, $\left| \Psi
_2\right\rangle $, $\cdots ,$ and $\left| \Psi _n\right\rangle $? We show
that it is indeed the case by directly proving the result.

{\it Theorem 4.} The states $\left| \Psi _1\right\rangle $, $\left| \Psi
_2\right\rangle $, $\cdots ,$ and $\left| \Psi _n\right\rangle $ can be
identified respectively with the efficiencies $\gamma _1,\gamma _2,\cdots ,$
and $\gamma _n$ if and only if the matrix $X^{\left( 1\right) }-\Gamma $ is
positive-semidefinite.

{\it Proof.} By definition, the identification is an $n+1$ outcome
measurement on the states $\left| \Psi _1\right\rangle $, $\left| \Psi
_2\right\rangle $, $\cdots ,$ and $\left| \Psi _n\right\rangle $. From the
general representation theorem for quantum operations [21], a general
measurement on system A can be represented by a unitary operation $U$ on the
composite system ABP, succeeded by a Von Neumann's type of measurement on
the probe P, where B indicates an ancillary system. In the idenfication
measurement, with the measurement outcome $P_i$ $\left( i=1,2,\cdots
,n\right) $, the state should be definitely $\left| \Psi _i\right\rangle $;
whereas with the $n+1$ outcome $P_{n+1}$, the state is not definitely
determined and the measurement fails. Hence the unitary operation $U$ on the
composite system ABP can be generally expressed as 
\begin{equation}
\label{13}U\left( \left| \Psi _i\right\rangle \left| \Psi _{BP}^{\left(
0\right) }\right\rangle \right) =\sqrt{\gamma _i}\left| \Psi _{AB}^{\left(
i\right) }\right\rangle \left| P_i\right\rangle +\sqrt{1-\gamma _i}\left|
\Phi _{AB}^{\left( i\right) }\right\rangle \left| P_{n+1}\right\rangle ,%
\text{ }\left( i=1,2,\cdots ,n\right) ,
\end{equation}
where $\gamma _i$ is the measurement efficiency with the input state $\left|
\Psi _i\right\rangle $, and $\left| P_1\right\rangle $, $\left|
P_2\right\rangle $, $\cdots $, and $\left| P_{n+1}\right\rangle $ are $n+1$
orthonormal states of the probe P. $\left| \Phi _{AB}^{\left( i\right)
}\right\rangle ,$ $\left| \Psi _{BP}^{\left( 0\right) }\right\rangle ,$ and $%
\left| \Psi _{AB}^{\left( i\right) }\right\rangle $ are normalized states of
the composite system AB, BP, and AB, respectively (not generally
orthogonal). Obviously, after the evolution (13), a Von Neumann's type of
measurement described by the projection operators $\left| P_i\right\rangle
\left\langle P_i\right| $ $\left( i=1,2,\cdots ,n+1\right) $ definitely
determines the input state with probability $\gamma _i$ of success.
Inter-inner-products of Eq. (13) yield the matrix equation 
\begin{equation}
\label{14}X^{\left( 1\right) }=\Gamma +\sqrt{I_n-\Gamma }\left[ <\Phi
_{AB}^{\left( i\right) }|\Phi _{AB}^{\left( j\right) }>\right] \sqrt{%
I_n-\Gamma }^{+}.
\end{equation}
Similar to the proof of theorem 2, semi-positivity of the matrix $X^{\left(
1\right) }-\Gamma $ thus becomes the necessary and sufficient condition for
identification of the states $\left| \Psi _1\right\rangle $, $\left| \Psi
_2\right\rangle $, $\cdots ,$ and $\left| \Psi _n\right\rangle $. This is
the content of theorem 4.

Theorem 4 determines the optimal measurement efficiencies. For example, if
there are three states $\left| \Psi _1\right\rangle ,$ $\left| \Psi
_2\right\rangle $ and $\left| \Psi _3\right\rangle $, and if $\left| \Psi
_1\right\rangle $ is orthogonal to $\left| \Psi _2\right\rangle $ and $%
\left| \Psi _3\right\rangle $, but $\left| \Psi _2\right\rangle $ is not
orthogonal to $\left| \Psi _3\right\rangle $, then the optimal efficiencies
are given by $\gamma _1=1,$ and $\frac{\gamma _2+\gamma _3}2\leq 1-\left|
<\Psi _2|\Psi _3>\right| $. The equality holds if and only if $\gamma
_2=\gamma _3$. This is essentially the result gained in Refs. [26-28], where
the identification of two nonorthogonal states is considered. For $n$
linearly independent and generally nonorthogonal states, the optimal
efficiencies $\gamma _1,\gamma _2,\cdots ,$ and $\gamma _n$ are obtainable
by solving a series of inequalities form the semi-positivity of the matrix $%
X^{\left( 1\right) }-\Gamma $.

In summary, we have shown that only linearly independent states can be
probabilistically cloned with non-zero probabilities of success. The best
possible cloning efficiencies are derived. We establish connection between
the probabilistic cloning and the identification measurement, and obtain the
optimal measurement efficiencies for $n$ linearly independent states.\\

{\bf Acknowledgment}

We thank Professor C. H. Bennett for critical comments and helpful
suggestions. This project was supported by the National Natural Science
Foundation of China.

\newpage\ 

\end{document}